\documentclass[twocolumn,twocolappendix]{aastex631}
\usepackage{amsmath}
\usepackage{bm}

\shorttitle{Magnetic Reconnection in BH Magnetospheres}
\shortauthors{Kimura, Toma, Noda, Hada}
\graphicspath{{./}{figures/}}

\begin{document}

\title{Magnetic Reconnection in Black-Hole Magnetospheres: \\
Lepton Loading into Jets, Superluminal Radio Blobs, and Multi-wavelength Flares}

\correspondingauthor{Shigeo S. Kimura}
\email{shigeo@astr.tohoku.ac.jp}

\author[0000-0003-2579-7266]{Shigeo S. Kimura}
\affiliation{Frontier Research Institute for Interdisciplinary Sciences, Tohoku University, Sendai 980-8578, Japan}
\affiliation{Astronomical Institute, Graduate School of Science, Tohoku University, Sendai 980-8578, Japan}

\author[0000-0002-7114-6010]{Kenji Toma}
\affiliation{Frontier Research Institute for Interdisciplinary Sciences, Tohoku University, Sendai 980-8578, Japan}
\affiliation{Astronomical Institute, Graduate School of Science, Tohoku University, Sendai 980-8578, Japan}

\author[0000-0001-6020-517X]{Hirofumi Noda}
\affiliation{Department of Earth and Space Science, Graduate School of Science, Osaka University, 1-1 Machikaneyama, Toyonaka, Osaka 560-0043, Japan}

\author[0000-0001-6906-772X]{Kazuhiro Hada}
\affiliation{Mizusawa VLBI Observatory, National Astronomical Observatory of Japan, 2-12 Hoshigaoka, Mizusawa, Oshu, Iwate 023-0861, Japan}
\affiliation{Department of Astronomical Science, The Graduate University for Advanced Studies (SOKENDAI), 2-21-1 Osawa, Mitaka, Tokyo 181-8588, Japan}



\begin{abstract}
Supermassive black holes in active galactic nuclei launch relativistic jets, as indicated by observed superluminal radio blobs. The energy source of these jets is widely discussed in the theoretical framework of Blandford-Znajek process, the electromagnetic energy extraction from rotating black holes (BHs), while formation mechanism of the radio blobs in the electromagnetically-dominated jets has been a long-standing problem.
Recent high-resolution magnetohydrodynamic simulations of magnetically arrested disks exhibited magnetic reconnection in a transient magnetically-dominated part of the equatorial disk near the BH horizon, which led to a promising scenario of efficient MeV gamma-ray production and subsequent electron-positron pair loading into BH magnetosphere.
We develop this scenario to build a theoretical framework on energetics, timescales and particle number density of the superluminal radio blobs and discuss observable signatures in other wavebands. We analytically show that the non-thermal electrons emit broadband photons from optical to multi-MeV bands. The electron-positron pairs produced in the magnetosphere are optically thick for synchrotron-self absorption, so that
the injected energy is stored in the plasma. The stored energy is enough to power the superluminal radio blobs observed in M87. This scenario predicts rather dim radio blobs around Sgr A*, which are consistent with no clear detection by current facilities.
In addition, this scenario inevitably produces strong X-ray flares in a short timescale, which will be detectable by future X-ray satellites. 
\end{abstract}



\section{Introduction} \label{sec:intro}
Relativistic jets are ubiquitously launched by black holes (BH), such as stellar mass BHs in X-ray binaries, central engines of gamma-ray bursts (GRBs), and supermassive BHs (SMBHs) in active galactic nuclei (AGNs). However, the production mechanisms of jets are still unclear. In the context of GRBs, the energy injection by neutrino annihilation is actively discussed \citep{KZ15a,2022arXiv220206480K}, whereas this mechanism does not work in X-ray binaries or AGNs. In these systems, the energy sources of jets are likely the spin energy of BHs threaded by magnetic fields, which can be extracted via Blandford-Znajek (BZ) process \citep{BZ77a,Kom04a,TT16a,2021PTEP.2021i3E03K}.

General relativistic magnetohydrodynamic (GR MHD) simulations of magnetized accretion disks show that the dilute polar region can form magnetosphere with $\sigma_B=B^2/(4\pi n_e m_e c^2) \gg 1$ around a BH, where the BZ process works \citep{2004ApJ...611..977M,TNM11a,2012MNRAS.426.3241N,TOK16a,2018ApJ...868..146N,2019ApJ...875L...5E,2022Univ....8...85M}. 
GRMHD simulations indicate that accretion flows can be classified into two modes \citep{2012MNRAS.426.3241N}: One is the standard and normal evolution (SANE) in which the magnetic fields are weak and turbulent, while the other is the magnetically arrested disk (MAD) in which magnetic fields are strong and mostly ordered. The electromagnetic power of jets in the MAD mode is higher than that for the SANE mode \citep{TNM11a,2019ApJ...875L...5E,2022NatAs...6..103C}, and thus, many believe that MADs are responsible for production of relativistic jets. This picture is observationally supported by the estimate of magnetic flux  using core-shift measurements of radio galaxies \citep{2014Natur.510..126Z} and the polarization map of M87 \citep{2021ApJ...910L..12E,2021ApJ...910L..13E}.

However, it has been a long-standing problem how BH magnetospheres produce superluminal radio blobs, which are observed as a remarkable indication of relativistic jets
\citep{1971ApJ...170..207C,1971Sci...173..225W,1984RvMP...56..255B,2019ARA&A..57..467B,2019Galax...8....1H}.
The previous GRMHD simulations and studies of non-thermal processes around BHs implied that the magnetosphere becomes too dilute to emit radiation ascribed to the superluminal radio blobs \citep{2011ApJ...730..123L,2011ApJ...735....9M,tt12,ktt14,2020ApJ...905..178K,2021ApJ...909..168K}.\footnote{Note that the wind or coronal regions of MADs outside the magnetosphere ($\sigma_B\lesssim 1$) does not have free energy sufficient for acceleration to relativistic speed \citep[e.g.,][]{SNP13a,2022arXiv220510531C}.} 
General relativistic (GR) particle-in-cell (PIC) simulations have not provided with any hints of sufficient $e^+ e^-$ pair production, either 
\citep{2018A&A...616A.184L,2018ApJ...863L..31C,2019PhRvL.122c5101P,2020ApJ...902...80K}.

Recently, \cite{2022ApJ...924L..32R} proposed a promising scenario of particle loading into jets.
In the MAD state, the strong and ordered magnetic fields can temporally halt the accretion process. Then, the accreting matter is accumulated outside a certain radius. This situation leads to development of magnetic Rayleigh-Taylor instability, which enables the matter to accrete to the BH with a spiral structure \citep{MTB12a,2019ApJ...874..168W}. Based on recent extremely high resolution GRMHD simulations, the magnetic fields above and below the equatorial plane push out the accreting matter in the low-density spiral, which triggers magnetic reconnection with $\sigma_B\gg1$ at the equatorial plane in BH magnetosphere  \citep{2022ApJ...924L..32R}. 
Such a relativistic magnetic reconnection accelerates non-thermal particles very efficiently \citep{2001ApJ...562L..63Z,2007ApJ...670..702Z,2014PhRvL.113o5005G,2014ApJ...783L..21S,2020PhPl...27h0501G}. The non-thermal electrons emit MeV gamma-rays via synchrotron radiation, which interact with each other and produce $e^+e^-$ pairs.  \cite{2022ApJ...924L..32R} pointed out that this mechanism can achieve the pair multiplicity of $10^8$ and they emit radio to optical photons by back-of-the-envelope estimates.

\begin{figure}
   \begin{center}
    \includegraphics[width=\linewidth,pagebox=cropbox]{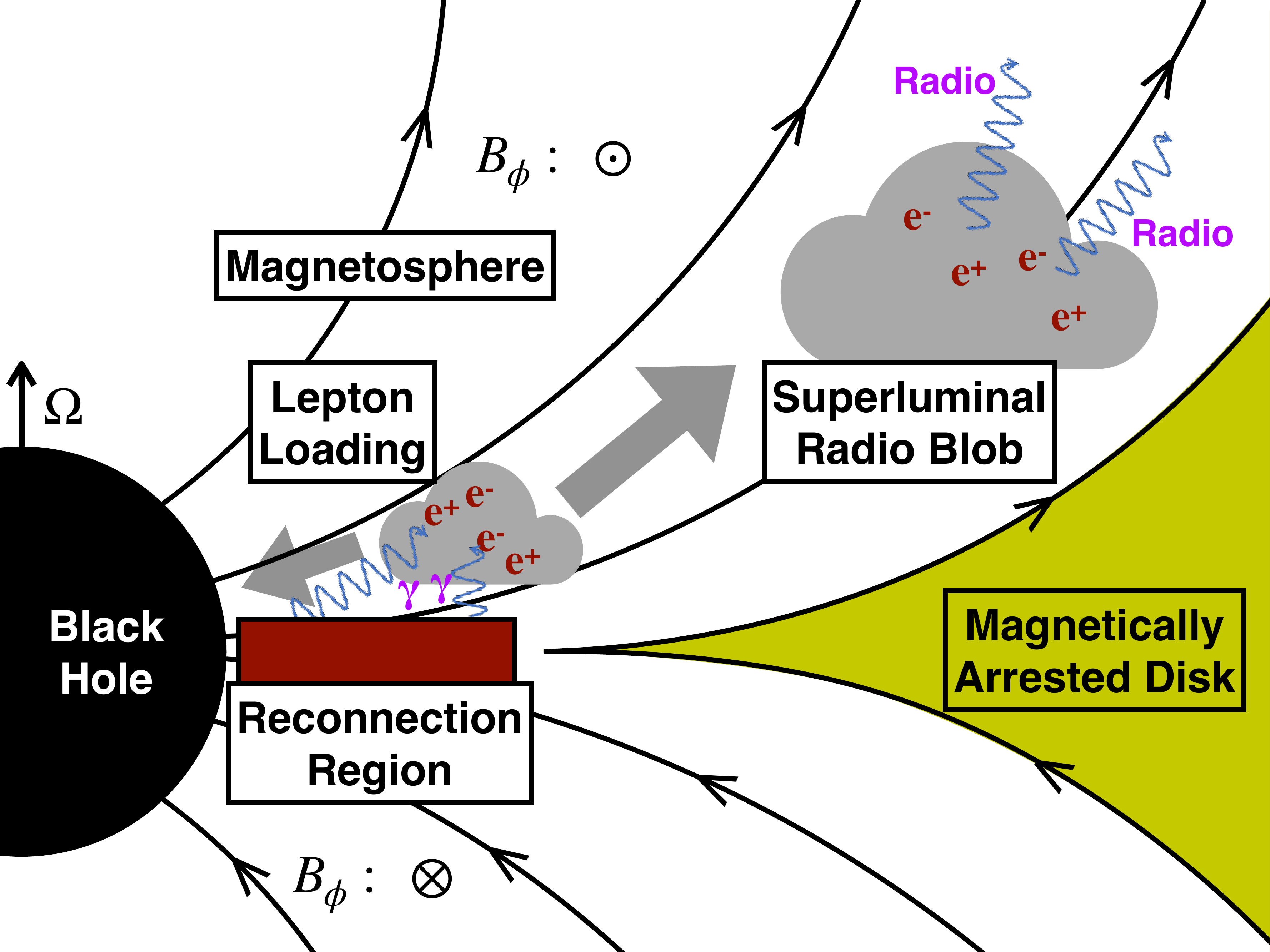}
    \caption{Schematic picture of our lepton loading scenario. 
    The magnetic field above and below the midplane of a MAD transiently pushes out the accreting matter in a non-axisymmetric manner, which triggers magnetic reconnection with $\sigma_B \gg 1$.
    The magnetic reconnection accelerates non-thermal electrons that emit luminous gamma-rays. 
    A fraction of these gamma-rays interact with each other, producing $e^+e^-$ pairs above and below the reconnection region. 
    The pair plasma is thermalized, and subsequently will be accelerated to relativistic speed.
    We can observe it as a superluminal radio blob after it becomes optically thin.}
    \label{fig:schematic}
   \end{center}
  \end{figure}

In this Letter, we further examine this lepton loading scenario and discuss observable signatures from the magnetic reconnection events. We analytically compute the synchrotron spectrum and pair production rate, and confirm that this process can load a large amount of $e^+e^-$ pairs. The bulk of the injected $e^+e^-$ pairs can be accelerated to a relativistic velocity, and their kinetic energy is sufficient to power a superluminal radio blob observed in radio galaxies (see Figure \ref{fig:schematic}). We also evaluate the detectability of X-ray flares during the reconnection event, and find that the future X-ray satellites will be able to detect the flares from Sgr A* and M87.  We use convention of $Q_x=Q/10^x$ throughout this Letter.

\section{Physical Conditions in Black-Hole Magnetosphere}

We consider a radio galaxy of SMBH mass $M=10^9 M_9 M_\odot$ with mass accretion rate of $\dot M=\dot mL_{\rm Edd}/c^2\simeq1.4\times10^{22}M_9\dot{m}_{-4}\rm~g~cm^{-2}$, where $c$ is the speed of light and $L_{\rm Edd}$ is the Eddington luminosity. The gravitational radius of the BH
is $r_g=GM/c^2\simeq1.5\times10^{14}M_9$ cm. We consider that the accretion flow is in the MAD state, and then
the magnetic field strength around the SMBH is estimated to be $ B_{\rm mad}=\sqrt{\dot Mc\Phi_{\rm mad}^2/(4\pi^2r_g^2)}\simeq1.1\times10^3M_9^{-1/2}\dot{m}_{-4}^{1/2}\Phi_{\rm mad,1.7}\rm~G$ \citep[e.g.,][]{YN14a}, where $\Phi_{\rm mad}\approx50\Phi_{\rm mad,1.7}$ is the saturated magnetic flux \citep{TNM11a,MTB12a,2012MNRAS.426.3241N,2019ApJ...874..168W}. 
The high-resolution GRMHD simulation with a BH spin parameter $a=0.9375$ suggests that magnetic reconnection occurs at a distance of $r_{\rm rec}\sim2r_g$ \citep{2022ApJ...924L..32R}. 
The value of $r_{\rm rec}$ could depend on $a$ or other parameters, but we fix $r_{\rm rec}=2r_g$ throughout this paper for simplicity. 
We estimate the reconnecting magnetic field strength to be (see Appendix \ref{sec:BrBphi})
\begin{align}
 B_{\rm rec}&\approx \sqrt{2}B_{\rm mad}\left(\frac{r_{\rm rec}}{r_g}\right)^{-2} \label{eq:Brec}\\
&\simeq3.9\times10^2M_9^{-1/2}\dot{m}_{-4}^{1/2}\Phi_{\rm rec,1.2}\rm~G,\nonumber
\end{align}
where $\Phi_{\rm rec}=\sqrt{2}\Phi_{\rm mad}(r_{\rm rec}/r_g)^{-2}$ is the effective magnetic flux at the reconnection region.

The magnetosphere will be formed around the SMBH. The minimum number density of the magnetosphere that can maintain the electric current for the BZ process is
\citep{1969ApJ...157..869G,2018A&A...616A.184L}
\begin{align}
 n_{\rm GJ}&=\frac{B_{\rm rec}\Omega_F}{2\pi ec} \approx\frac{B_{\rm rec}}{8\pi e r_g}\\
 &\simeq 2.2\times10^{-4}M_9^{-3/2}\dot{m}_{-4}^{1/2}\Phi_{\rm rec,1.2}\rm~cm^{-3},\nonumber
\end{align}
where $\Omega_F \approx ac/(4r_g)$ is the field line angular velocity \citep{2010ApJ...711...50T,2014ApJ...788..186N,2021ApJ...911...34O,2022JCAP...07..032C} and we assume the BH spin parameter as $a\sim 1$. For the magnetosphere which consists of $e^+ e^-$ pair plasma with the density $n_{\rm GJ}$,
the magnetization parameter is 
\begin{align}
 \sigma_{B,\rm GJ}=\frac{B_{\rm rec}^2}{4\pi n_{\rm GJ}m_ec^2}\approx 6.8\times10^{13}M_9^{1/2}\dot{m}_{-4}^{1/2}\Phi_{\rm rec,1.2}.\label{eq:sigBgj}
\end{align}
This value should be regarded as an upper limit, because the number density of the magnetosphere can be higher than $n_{\rm GJ}$. Various mechanisms of particle injection into BH magnetosphere have been proposed (see Appendix \ref{sec:massloading}), which can lead to multiplicity of $\kappa_\pm \equiv n/n_{\rm GJ}\sim1-10^3$. This results in the magnetization parameter of $\sigma_B\gtrsim10^{10}$.

As for the reconnection region, where the accreting gas is being pushed out, GRMHD simulations imply that the possible range is $100\ll \sigma_B\lesssim\sigma_{B,\rm GJ}$.
In the following discussion, we assume $\sigma_B$ is higher than the value of $\gamma_{e,{\rm max}}$ introduced below (Equation \eqref{eq:gamemax}).

\section{Magnetic Reconnection in BH magnetosphere}\label{sec:flare}

The relativistic magnetic reconnection accelerates non-thermal particles very efficiently (see Appendix \ref{sec:acceleration} for discussion). If we ignore the effect of cooling, almost all the particles are accelerated up to the energy of $\gamma_e\sim\sigma_B$  \citep{2020PhPl...27h0501G}. At the initial stage, the particles are accelerated by the reconnection electric field in a timescale of $t_{\rm acc}\approx \gamma_e m_e c/(eB_{\rm rec}\beta_{\rm rec})$, where $\beta_{\rm rec}\sim0.1$ is the reconnection velocity in the kinetic domain \citep[e.g.,][]{2020PhPl...27h0501G}.
The accelerated electrons cool by the synchrotron emission with a timescale of $t_{\rm syn}=6\pi m_ec/(\sigma_TB_{\rm rec}^2\gamma_e)$. Equating the acceleration time and the cooling time, we obtain the maximum Lorentz factor as
\begin{align}
 \gamma_{e,\rm max}\approx\sqrt{\frac{6\pi e \beta_{\rm rec}}{\sigma_TB_{\rm rec}}}\simeq1.9\times10^6M_9^{1/4}\dot{m}_{-4}^{-1/4}\Phi_{\rm rec,1.2}^{-1/2}\beta_{\rm rec,-1}^{1/2}.\label{eq:gamemax}
\end{align}
The synchrotron frequency for these electrons are
\begin{align}
 E_{\gamma,\rm max}&=\frac{heB_{\rm rec}\gamma_{e,\rm max}^2}{2\pi m_ec} 
 =\frac{9m_e c^2}{4\alpha_f}\beta_{\rm rec}\nonumber\\
 &\simeq 16 \beta_{\rm rec,-1}\rm~MeV,\label{eq:Egammax}
\end{align}
where $h$ is the Planck constant and $\alpha_f$ is the fine structure constant.
The value of $E_{\gamma,\rm max}$ depends only on $\beta_{\rm rec}$, whose value is almost constant according to recent PIC simulations.\footnote{The value of $\beta_{\rm rec}$ is almost constant in time for each PIC simulation, and the variation among the different PIC simulations are also small \citep[e.g.,][]{2020PhPl...27h0501G}} 

The energy release rate by the reconnection event is estimated to be
\begin{align}
 L_{\rm rec}&\approx 2 l_{\rm rec}^2\frac{B_{\rm rec}^2}{8\pi}\beta_{\rm rec} c\label{eq:Lrec}\\
&\simeq7.9\times10^{41}M_9\dot{m}_{-4}f_l^2\beta_{\rm rec,-1}\Phi_{\rm rec,1.2}^2\rm~erg~s^{-1},\nonumber
\end{align}
where $l_{\rm rec}=f_{l}r_g$ is the length scale of the reconnection region and $f_{l}$ is a parameter.  
The duration of the reconnection event can be estimated to be
\begin{align}
 T_{\rm dur}\approx \frac{h_{\rm rec}}{2\beta_{\rm rec}c}\simeq2.5\times10^4M_9\xi_{hl,-0.3}\beta_{\rm rec,-1}^{-1}\rm~sec,
\end{align}
where $h_{\rm rec}=\xi_{hl}l_{\rm rec}$ is the thickness of the anti-parallel magnetic fields that will reconnect during the reconnection event, and $\xi_{hl}$ is a parameter. We consider that large-scale dynamics should determine $h_{\rm rec}$ and $l_{\rm rec}$, whereas microscopic plasma properties should determine the thickness of the reconnection layer where magnetic fields are dissipating. Since the system is in the fast cooling regime, all the released energy is converted to photons by synchrotron emission. 
We approximately suppose that the reconnection event injects mono-energetic particles at $\gamma_e\sim\gamma_{e,\rm max}$ and these particles are confined in magnetic islands. Then, we can write the broadband synchrotron photon spectrum as \citep[e.g.,][]{SPN98a}
\begin{align}
 E_\gamma L_{E_\gamma}\approx L_{\rm rec}\left(\frac{E_\gamma}{E_{\gamma,\rm max}}\right)^{1/2}.\label{eq:ELE}
\end{align}
The power-law spectrum continues to the cooling frequency that is typically in radio bands. We should note that this photon spectrum is different from that obtained in 2D PIC simulations by \cite{2013ApJ...770..147C,2019ApJ...877...53H}, whose spectrum is extended beyond the burn-off limit ($\sim160$ MeV; see Appendix \ref{sec:acceleration} for discussion).

  \begin{figure}
   \begin{center}
    \includegraphics[width=\linewidth,pagebox=cropbox]{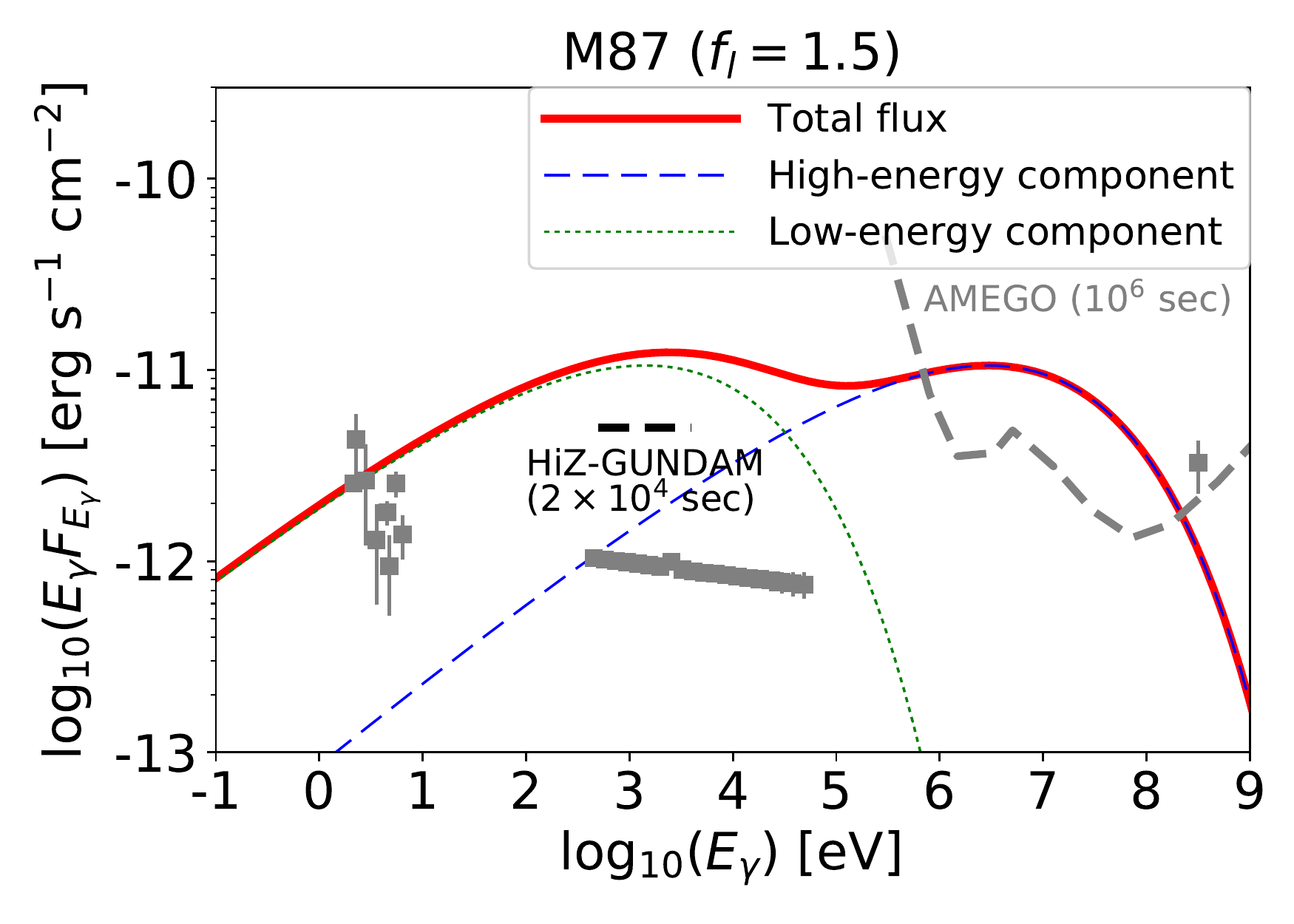}
    \caption{Photon spectrum of a reconnection-driven flare from M87. Parameters are $M=6.3\times10^9M_\odot$, $\dot{m}=5\times10^{-5}$, $f_{l}=1.5$, and $\xi_{hl}=0.5$. The blue-dashed, green-dotted, and red-solid lines are for the high-energy flaring state, the low-energy flaring state, and sum of them, respectively. The data points are obtained from Table A8 in \cite{2021ApJ...911L..11E}, which is in the quiescent state. Our model predicts flares of $\sim10$ times higher luminosity. The black- and grey-dotted lines are sensitivity curves for HiZ-GUNDAM ($2\times10^4$ sec: \citealt{2020SPIE11444E..2ZY}) and AMEGO ($10^6$ sec: \citealt{2019BAAS...51g.245M}), respectively. }
    \label{fig:flare}
   \end{center}
  \end{figure}

The photons in MeV energies interact with each other and create $e^+e^-$ pairs. Using the photon spectrum of Equation (\ref{eq:ELE}), the optical depth for $\gamma\gamma$ interaction is estimated to be 
\begin{align}
\tau_{\gamma\gamma}\approx n_{\gamma_2}\sigma_{\gamma\gamma}l_{\rm rec}\simeq 1.2\times10^{-3} \dot{m}_{-4}f_{l}\beta_{\rm rec,-1}\Phi_{\rm rec,1.2}^2,\label{eq:taugam}
\end{align}
where $n_{\gamma_2}$ is the number density of target photons (see Appendix \ref{sec:multiplicity}) and  $\sigma_{\gamma\gamma}\approx f_{\gamma\gamma}\sigma_T$ is the approximated cross section for two-photon interaction, $\sigma_T$ is the Thomson cross section, and $f_{\gamma\gamma}\sim0.2$. Thus, vast majority of the gamma-rays escape from the system.
Nevertheless,
as shown in Appendix \ref{sec:multiplicity}, we find a high pair multiplicity as
\begin{align}
 \kappa_\pm &= \frac{n_\pm}{n_{\rm GJ}} \approx
 \frac{4}{3}\alpha_f \tau_{\gamma\gamma} \sigma_{B,{\rm GJ}} \label{eq:kappapm}\\
&\simeq7.7\times10^8 M_9^{1/2}\dot{m}_{-4}^{3/2}f_{l}\Phi_{\rm rec,1.2}^3\beta_{\rm rec,-1} \nonumber
\end{align}
This mechanism can achieve several orders of magnitude higher values of multiplicity than those by the previously proposed scenarios. This is because the relativistic reconnection event can efficiently convert the magnetic energy into gamma-rays of MeV energies in the compact region.

The leptons loaded into the pre-reconnection plasma change the magnetization parameter there, which affects the photon spectra and the pair production rate. In our fiducial parameters, the magnetization parameter after the lepton loading is estimated to be
\begin{equation}
\sigma_B = \frac{\sigma_{B,\rm GJ}}{\kappa_\pm}\simeq 8.7\times10^4 \dot{m}_{-4}^{-1}f_{l}^{-1}\Phi_{\rm rec,1.2}^{-2}\beta_{\rm rec,-1}^{-1} \label{eq:sigB}
\end{equation}
Then, $e^+e^-$ pairs are accelerated up to $\gamma_e\sim \sigma_B$, and the flare spectrum has a cutoff at
\begin{align}
 E_{\gamma,\rm syn}&=\frac{heB_{\rm rec}\sigma_B^2}{2\pi m_ec}\\
&\simeq 35M_9^{-1/2}\dot{m}_{-4}^{-3/2}f_{l}^{-2}\Phi_{\rm rec,1.2}^{-3}\beta_{\rm rec,-1}^{-2}\rm~keV.\nonumber
\end{align}
Thus, the flare peak energy will significantly decrease after the lepton loading, which occurs around light-crossing timescale, $l_{\rm rec}/c$. 
We call the spectral state with $\sigma_B$ as `low-energy flaring state' and that with $\sigma_{B,{\rm GJ}}$ as `high-energy flaring state'.
Since the low-energy flaring state cannot produce MeV photons efficiently, the $e^+e^-$ pairs are not produced.

The injected $e^+e^-$ pairs in the pre-reconnection plasma rotate with an angular velocity of $\Omega_{\rm high}\sim\Omega_F\sim c/(4r_g)$. On the other hand, according to GRMHD simulations \citep[e.g.,][]{2012MNRAS.426.3241N,MTB12a}, the plasma at midplane will have the angular velocity of $\Omega_{\rm mid}\sim \Omega_{\rm mad}\ll \Omega_K$, where $\Omega_{\rm mad}$ is the angular velocity in the MAD and $\Omega_K\simeq0.26c/r_g$ at $r\simeq 2r_g$ is the Keplerian angular velocity. Thus, $\Omega_{\rm mid}\ll \Omega_{\rm high}$ should be satisfied, and the injected $e^+e^-$ pairs will escape in a streaming timescale of $\sim 2l_{\rm rec}/c$. This timescale is shorter than the duration of the reconnection event, $T_{\rm dur}$. Hence, after the streaming escape timescale, the reconnection event will become able to accelerate electrons up to $\gamma_{e,\rm max}$ again, leading to efficient MeV gamma-ray production. These gamma-rays will produce the copious $e^+e^-$ pairs again, and thus, we expect oscillation between the low- and high-energy flaring states. The timescale of pair production and streaming escape is comparable, and thus, we consider that the duration for each state should be similar.

\begin{table*}
\begin{center}
\caption{Resulting physical quantities in M87 and Sgr A*. We use $f_l=1.5$ for M87, $f_l=0.6$ for Sgr A*, and $\xi_{hl}=0.5$ for both objects. We use $M=6.3\times10^9 M_\odot$, $\dot{m}=5\times10^{-5}$, and $d_L=17$ Mpc for M87 \citep{2019ApJ...875L...6E,2020ApJ...905..178K} and $M=4.0\times10^6 M_\odot$, $\dot{m}=6\times10^{-7}$, and $d_L=8$ kpc for Sgr A* \citep{EventHorizonTelescope:2022xnr,2022arXiv220509565K}. } \label{tab:quant}
 \begin{tabular}{|c|ccccccc|}
\hline
 Name & $B_{\rm rec}$ & $\log_{10}(n_{\rm GJ})$ & $\log_{10}(\sigma_{B,\rm GJ})$ & $\log_{10}(L_{\rm rec})$ & $\log_{10}(T_{\rm dur})$ & $\log_{10}(\kappa_\pm)$ & $F_{\rm radio}$ (43 GHz) \\
 & [G] & [cm$^{-3}$] & & [erg s$^{-1}$] & [s] & & [mJy]\\
\hline
M87 & 1.1$\times10^2$ & -5.0 & 14.1 & 42.8 & 5.4 & 9.0 & 12  \\
Sgr A* & 4.8$\times10^2$ & -1.2 & 11.5 & 36.8 & 1.8 & 4.1 & 0.07  \\
\hline
\end{tabular}
\end{center}
\end{table*}

To discuss observational signature, we numerically calculate the photon spectra by a reconnection event. We solve the transport equation for the non-thermal leptons accelerated by the reconnection, and calculate the synchrotron photon spectrum \citep[see, e.g.,][for technical details]{2020ApJ...904..188K}. The BH mass and mass accretion rate for M87 can be estimated by other studies \citep[e.g.,][]{2019ApJ...875L...1E,2020ApJ...905..178K}, and  we can regard $\beta_{\rm rec}$ and $\Phi_{\rm rec}$ as constants, 
based on relativistic PIC simulations \citep{2020PhPl...27h0501G} and GRMHD simulations \citep{2012MNRAS.426.3241N,2022ApJ...924L..32R}, respectively.
Then, $f_{l}$ and $\xi_{hl}$ are the only free parameters in our scenario. We choose $f_{l}=1.5$ and $\xi_{hl}=0.5$ so that the pair plasma loaded in the magnetosphere can power the observed radio blobs (see Section \ref{sec:blob}). We cannot explain the power of radio blobs with a smaller value of $f_{l}$.

Figure \ref{fig:flare} shows the resulting spectra with a parameter set for M87, where we average over the flux from each state. Our model predicts a broadband flare whose flux in X-ray bands is about 10 times higher than that for the quiescent state. The core of M87 exhibited such a high flux state in the past \citep[e.g.,][]{2012ApJ...746..151A}.
Based on the high-resolution GRMHD simulation \citep{2022ApJ...924L..32R}, the magnetic reconnection repeatedly occurs in a timescale of $T_{\rm int}\sim 2000 r_g/c\sim2$ yr. On the other hand, the duration of flare is very short, $T_{\rm dur}\sim3$ days (see Table \ref{tab:quant}). Thus, we need to monitor M87 for years with a cadence of less than a few days. We cannot occupy high-sensitivity pointing facilities, e.g., Chandra, for such a long time. Thus, we need to rely on all-sky X-ray monitors, such as Swift-BAT \citep{SwiftBAT05a} and MAXI \citep{2009PASJ...61..999M}. Although the predicted flux is much lower than the sensitivities for current X-ray monitors, future X-ray monitors, such as HiZ-GUNDAM \citep{2020SPIE11444E..2ZY} and Einstein Probe \citep{2015arXiv150607735Y}, can detect the X-ray flares.
The timescale of the oscillation between the low- and high-energy flaring states is $2l_{\rm rec}/c\simeq$ 93 ksec, which is longer than the integration time required for detection by HiZ-GUNDAM (see Figure \ref{fig:flare}). 
Also, future X-ray satellite networks might enable the sensitive pointing X-ray telescopes to follow up the flare event within an hour.
Then, we will be able to observe the oscillation between the low- and high-energy flaring states, which will enable us to determine the values of $f_l$ and $\xi_{hl}$. Therefore, our scenario will be robustly tested by future X-ray observations.

\section{Origin of Superluminal radio blobs}\label{sec:blob}

The $e^+e^-$ pairs injected in the magnetosphere can lose energies by synchrotron radiation, but they may be heated up by synchrotron-self absorption (SSA) process. As shown in Appendix \ref{sec:SSAheating}, the leptons are easily thermalized in typical radio galaxies.
At the time of lepton loading, we can write the spectrum of the $e^+e^-$ pairs as a power-law with index of $s\sim1$, i.e., $N_{E_\pm}\propto E_\pm^{-1}$, where $E_\pm$ is the energy of the $e^+e^-$ pairs. This is a good approximation if the parent photon spectrum is a power-law (see Appendix \ref{sec:multiplicity}).
Considering the number and energy conservations during the thermalization, we obtain
\begin{align}
k_BT_\pm\approx \frac{E_{\pm,\rm max}}{3\ln(E_{\pm,\rm max}/E_{\pm,\rm min})}\sim 0.96\beta_{\rm rec,-1} \rm~MeV,
\end{align}
where $E_{\pm,\rm max}\approx E_{\gamma,\rm max}/2$ and $E_{\pm,\rm min}\sim m_ec^2$ are the maximum and minimum energies of the $e^+e^-$ pairs, respectively.
 The mean energy of the thermalized $e^+e^-$ pairs is $3k_BT_\pm\simeq2.9\beta_{\rm rec,-1}$ MeV, and thus, we can regard this as a relativistic plasma. Such a hot gas can accelerate to relativistic speed by its adiabatic expansion.

The thermalized pairs do not cool via the synchrotron process in the initial phase.
We can write the synchrotron peak frequency for thermal plasma as $\nu_{\rm blb,syn}\simeq 9\theta_\pm^2eB_{\rm blb}/(2\pi m_ec)$ and the emissivity at $\nu =\nu_{\rm blb,syn}$ as $j_\nu = 3\sigma_T c \theta_{\pm}^2 B_{\rm blb}^2 n_{\rm blb}/(2\pi \nu_{\rm blb,syn})$, where $B_{\rm blb}$ and $n_{\rm blb}$ are the magnetic field strength and the number density of the blob and $\theta_\pm=k_BT_\pm/(m_ec^2)$. 
The optical depth at the synchrotron peak can roughly be estimated to be
\begin{align}
 \tau_{\rm ssa,blb}&\approx \frac{j_\nu c^2R}{2\nu_{\rm blb,syn}^2k_BT_\pm}\approx 
 \frac{16\pi^3}{3^6}\frac{e}{B_{\rm blb}\theta_{\rm blb}^5}n_{\rm blb} R \\
&\simeq 9.0\times10^5 M_9^{1/2}\dot{m}_{-4}^{3/2}f_{l}^2\Phi_{\rm rec,1.2}^3\beta_{\rm rec,-1}^{-4}\nonumber
\end{align}
where we use the values immediately after the lepton injection with $R \sim f_l r_g$ in the estimate.
The thermal plasma cannot cool by the synchrotron process while $\tau_{\rm ssa,blb} > 1$.

The total energy of the lepton plasma during the reconnection event is estimated to be 
\begin{align}
 \mathcal{E}_{\rm blb}&\approx \dot{N}_{\gamma\gamma}T_{\rm dur}E_{\pm,\rm max}\\
 &\simeq 1.1\times10^{43} M_9^2\dot{m}_{-4}^2f_{l}^4\xi_{hl,-0.3}\beta_{\rm rec,-1}\Phi_{\rm rec,1.2}^2 \rm~erg,\nonumber
\end{align}
Theoretically, it is unknown how such a non-steady, non-axisymmetric MHD flow accelerates, whereas steady, axisymmetric MHD acceleration of hot flows was extensively studied with possible cooling effects \citep{2003ApJ...596.1080V,2009MNRAS.394.1182K,2020MNRAS.494..338T}.
According to VLBI observations, the blob velocities are widely distributed for $\beta \lesssim 0.7$ at $R \sim 100-1000r_g$
\citep{2019ApJ...887..147P,2018ApJ...868..146N}.
The injected energy can be dissipated by internal shocks \citep{2012MNRAS.422..326K} or plasma kinetic instabilities \citep[e.g.,][]{2021ApJ...907L..44S,2022ApJ...928...62K}, which leads to non-thermal lepton acceleration. These leptons emit radio signals where $\tau_{\rm ssa,blb} < 1$, and we observe it as a radio blob. 
The radio luminosity and flux from the blob can be estimated to be
\begin{align}
 L_{\rm radio}&\approx \frac{\mathcal{E}_{\rm blb}}{R_{\rm dis}/c}\simeq2.3\times10^{37}M_9\dot{m}_{-4}^2f_{l}^4\label{eq:Lradio}\\
&\times \xi_{hl,-0.3}\beta_{\rm rec,-1}\Phi_{\rm rec,1.2}^2\mathcal{R}_2^{-1}\rm~erg~s^{-1},\nonumber \\
 F_{\rm radio}&\sim \frac{L_{\rm radio}}{4\pi d_L^2 \nu_{\rm blob}}
\end{align}
where $R_{\rm dis}=\mathcal{R}r_g$ is the dissipation radius, $\mathcal{R}$ is a parameter, and $\nu_{\rm blob}$ is the radio frequency. The value of $F_{\rm radio}$ for M87 is shown in Table \ref{tab:quant}. VLBI observations detect radio blobs of $F_{\rm radio}\sim10$ mJy at $R\sim100-1000 r_g$ in M87  \citep{2013ApJ...775...70H,2018ApJ...855..128W,2019ApJ...887..147P}, which is in rough agreement with our prediction. 
The injection and expansion of pairs are concentrated on the edge of the magnetosphere, which may also be compatible with the limb-brightened feature of the jet \citep[cf.,][]{2018A&A...616A.188K,2018ApJ...868...82T}.

\section{Application to Sgr A*} \label{sec:sgrA}

  \begin{figure}
   \begin{center}
    \includegraphics[width=\linewidth,pagebox=cropbox]{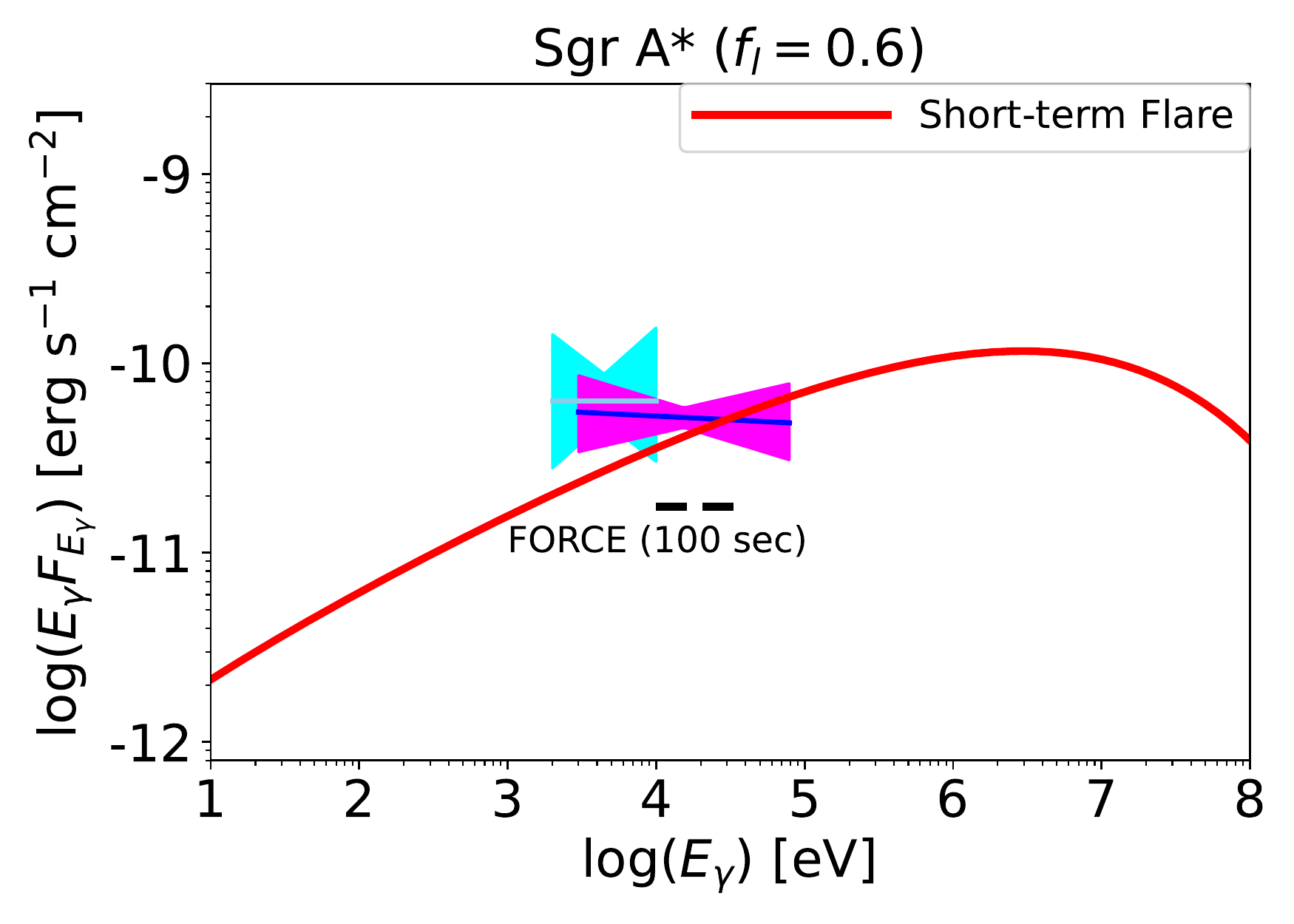}
    \caption{Photon spectrum of a reconnection-driven flare from Sgr A* (red-solid line). Parameters are $M=4.0\times10^6M_\odot$, $\dot{m}=6\times10^{-7}$, $f_l=0.6$, and $\xi_{hl}=0.5$. The X-ray flare data (cyan and magenta regions) are taken from \cite{2012ApJ...759...95N} and \cite{2014ApJ...786...46B}, respectively. The black-dashed line is sensitivity curve for FORCE (100 sec: \citealt{2018SPIE10699E..2DN}).}
    \label{fig:SgrA}
   \end{center}
  \end{figure}

Recent studies suggest that Sgr A* likely has a MAD. Large-scale GRMHD simulations revealed that accretion by the Wolf-Rayet wind leads to a formation of a MAD around Sgr A* \citep{2020ApJ...896L...6R}, which is consistent with the variability of Sgr A* \citep{2022arXiv220406170M}. Sgr A* is known as a very low accretor, which likely leads to a MAD state owing to a rapid advection of large scale magnetic fields \citep{2011ApJ...737...94C,2021ApJ...915...31K}. Recent radio imaging of the horizon scale also favors a relatively face-on MAD state in Sgr A* \citep{EventHorizonTelescope:2022urf}, although the models with SANE-mode RIAFs are not ruled out.

We apply our scenario to Sgr A*, tabulate resulting physical quantities in Table \ref{tab:quant}, 
and show the resulting flare spectrum in Figure \ref{fig:SgrA}. Here, we use $f_l=0.6$ and $\xi_{hl}=0.5$ so that the resulting X-ray flux is consistent with the observed X-ray flux for Sgr A* flares \citep[e.g.,][]{2012ApJ...759...95N,2013ApJ...774...42N,2014ApJ...786...46B}.\footnote{The value of $f_l$ for Sgr A* can be different from that for M87, because the large-scale magnetic field configuration may be different. Sgr A* may accrete mass from the Wolf-Rayet winds \citep{2020ApJ...896L...6R}, whereas M87 is expected to accrete mass from the interstellar medium.} 
Owing to the low lepton loading rate, Sgr A* can maintain a high magnetization even after the lepton loading (see Equation (\ref{eq:sigB})),
which causes the reconnection flare for Sgr A* to exhibit only the high-energy state. 
The near infrared (NIR) flux by our scenario is comparable to that observed in the quiescent state.
The duration of the reconnection-driven flare is around $T_{\rm dur}\sim10^2$ sec, which is shorter than the time bin for the current flare analyses for Sgr A*, $\Delta t\sim300$ sec \citep[e.g.,][]{2013ApJ...774...42N}. \cite{2012ApJ...759...95N,2014ApJ...786...46B} reported substructure of $\sim100$-sec duration in the lightcurve of bright X-ray flares, which may be caused by the reconnection-driven flare. We need more statistics to confirm this feature, and the future hard X-ray satellite, FORCE \citep{2018SPIE10699E..2DN}, will be able to observe a rapid variability with a high statistics. 

\begin{figure}
   \begin{center}
    \includegraphics[width=\linewidth,pagebox=cropbox]{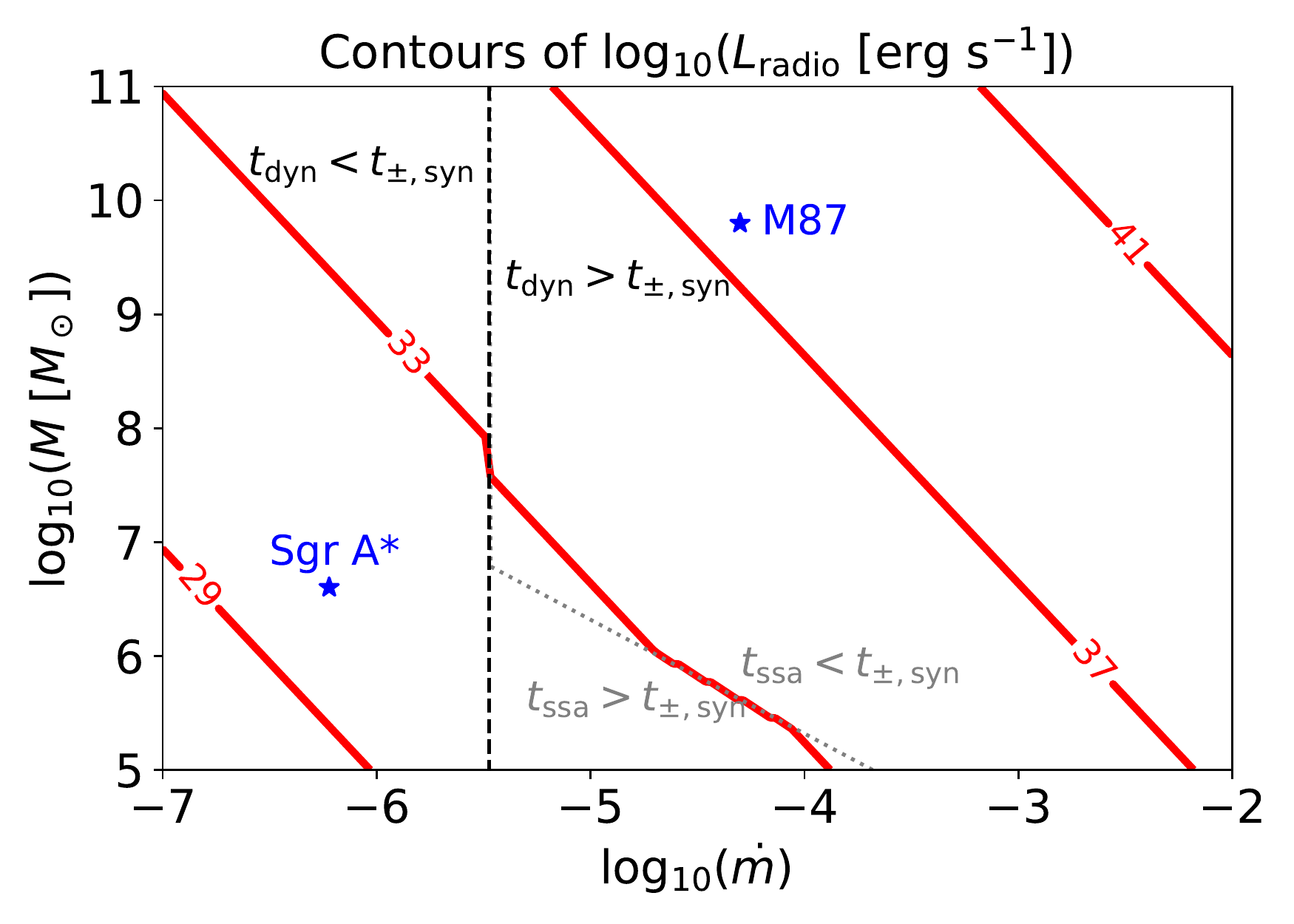}
    \caption{Contours of radio luminosity as a function of $M$ and $\dot{m}$ with $f_l=1.0$ and $\xi_{hl}=0.5$. The red-lines show the contours of $\log_{10}(L_{\rm radio})$ in cgs unit. The gray-dotted line shows $t_{\rm ssa}=t_{\pm,\rm syn}$, and black-dashed line shows $t_{\rm dyn}=t_{\pm,\rm syn}$. The blue stars show the parameters for Sgr A* and M87.}
    \label{fig:mdot_M}
   \end{center}
  \end{figure}

Our model predicts that Sgr A* may have radio blobs at $R\sim100-1000r_g$, but the radio flux from Sgr A* is too faint to clearly detect with current facilities. 
Figure \ref{fig:mdot_M} shows $L_{\rm radio}$ as a function of $M$ and $\dot{m}$, where we see $L_{\rm radio}$ strongly depends on $M$ and $\dot{m}$ (see Appendix \ref{sec:SSAheating} for the method for estimating $\mathcal{E}_{\rm blb}$; see also Equation (\ref{eq:Lradio})). 
We find that $F_{\rm radio}$ for Sgr A* is $\sim0.1$ mJy, which may be challenging to detect. VLBI images on Sgr A* are likely affected by the refractive scattering effect in the foreground plasma, which effectively adds $1-10$ mJy noises to the data \citep{2022ApJ...926..108C}.  \cite{2019A&A...621A.119B} reported a possible blob-like feature in their 86 GHz VLBI Sgr A* image with a flux density of $\sim10$ mJy. However, it is again still unclear whether this sub-structure is intrinsic to Sgr A* or caused by the scattering effect. Higher sensitivity and dedicated monitoring observations would be demanded to firmly detect radio blobs around Sgr A*.

The observed duration of Sgr A* X-ray flares is $\sim10^3-10^4$ sec, and the observed flare rate is  about $\sim0.1-0.3$ day$^{-1}$ for the ones as bright as in Figure \ref{fig:SgrA} \citep{2013ApJ...774...42N,2013ApJ...769..155D}. Our scenario predicts a shorter flare duration of $T_{\rm dur}\sim10^2$ sec and a higher flare rate of $\sim1-2$ day$^{-1}$.
Thus, our scenario cannot explain the typical timescales of the observed X-ray flares.\footnote{ We should note that the energy of the individual $e^+e^-$ particles injected by $\gamma\gamma$ pair production is too low to emit X-rays, and thus, the newly injected $e^+e^-$ pairs cannot explain X-ray flares in Sgr A*, either.}
Nevertheless, our scenario could reconcile with the observed flare properties. The magnetic reconnection in magnetosphere should produce bi-polar outflows, one of which collide with the dense plasma in a MAD \citep{2022ApJ...924L..32R}. This collision triggers another magnetic reconnection event at $R\sim 5-30r_g$, which will produce a flare with longer duration \citep{2020MNRAS.497.4999D,2021MNRAS.502.2023P,2022MNRAS.511.3536S}. Also, in Sgr A*, the wind-fed accretion leads to a weaker variability \citep{2022arXiv220406170M}. This could be caused by a weaker turbulent field in the wind-fed accretion flow, which implies that development of Rayleigh-Taylor instability likely takes more time than typical AGN that accretes turbulent fields from the outer accretion disk. This would lead to a longer interval between the reconnection events, which alleviates the tension of the flaring rate.

Sgr A* exhibits strong flares in the NIR band, and they may be powered by thermal electrons heated by the magnetic reconnection in the magnetically dominated regions in the MAD \citep{2020MNRAS.497.4999D,2022MNRAS.511.3536S}. 
In our scenario, a large number of $e^+e^-$ pairs are injected in the magnetosphere, and they emit sub-mm emission via synchrotron at $R\sim1-10r_g$, as the $e^+e^-$ plasma in Sgr A* is SSA thin. 
They would emit NIR photons if they were additionally heated by some mechanism. However, the number density of the $e^+e^-$ pairs in the magnetosphere ($n_{\rm GJ}\sim0.06\rm~cm^{-3}$) is much lower than that in the surrounding accretion flow.
Even after the pair injection, the magnetization parameter is $\sigma_B>10^7$, which is much higher than that investigated by GRMHD simulations \citep[$\sigma_B\sim10^2$, see e.g.,][]{2022MNRAS.511.3536S}. Thus, it is unlikely that the injected $e^+e^-$ pairs power the NIR flares in Sgr A*.
\cite{2018A&A...618L..10G} reported orbital motions of bright spots in NIR flares. These may be caused by outflows interacting with the surrounding accretion flow \citep{2020MNRAS.497.2385M}. The reconnection in BH magnetosphere can potentially trigger the NIR flares by providing outflows.

\section{Summary}

We have examined magnetic reconnection in BH magnetosphere and subsequent non-thermal particle production as a lepton loading mechanism to AGN jets. Extremely high-resolution GRMHD simulation by \cite{2022ApJ...924L..32R} suggested that magnetic reconnection repeatedly occurs in a magnetically-dominated part of the equatorial disk near the BH horizon. The magnetic energy is efficiently converted to the energy of non-thermal electrons, which subsequently cool by MeV gamma-ray flares.
These gamma-rays interact with each other and produce copious $e^+e^-$ pairs. We have evaluated the amount of $e^+e^-$ pairs and found that the plasma density around the reconnection region can be $\sim10^9$ times higher than the Goldreich-Julian density for a typical radio galaxy, which is sufficient for explaining signals from radio jets. 
The injected $e^+e^-$ pairs are optically thick for SSA process, and thus, the $e^+e^-$ plasma stores all the injected energy that is enough to power the observed superluminal radio blobs in M87. 
Our scenario predicts that radio blobs may exist around Sgr A* if it has a MAD, but they are faint and consistent with no clear detection by current facilities.
In our scenario, the reconnection event inevitably produces X-ray signals detectable by future X-ray satellites, such as HiZ-GUNDAM and FORCE, which will provide a concrete test for our scenario.

\vspace{10pt}
We thank K. Kashiyama, S. Kisaka, Y. Kojima, S. J. Tanaka, K. Tomida, and S. Tomita for fruitful discussions.
This work is partly supported by KAKENHI No. 22K14028 (S.S.K.), No. 18H01245 (K.T.), No. 19K21884, 20H01941, 20H01947 (H.N.), No. 21H01137, 22H00157 (K.H.). This work is supported by the Tohoku Initiative for Fostering Global Researchers for Interdisciplinary Sciences (TI-FRIS) of MEXT's Strategic Professional Development Program for Young Researchers.

\appendix

\section{Magnetic field structure in BH magnetosphere}\label{sec:BrBphi}

We estimate the structure of magnetic field in BZ process. We use the Boyer-Lindquist coordinates $(t, \varphi, r, \theta)$ and the unit of $GM=1$ and $c=1$ (and thus $r_g = 1$) in this Appendix. The metric of Kerr space-time can be written as
\begin{equation}
    ds^2 = -\alpha^2 dt^2 + \gamma_{ij}(\beta^i dt + dx^i)(\beta^j dt + dx^j),
\end{equation}
where the non-zero components are $\alpha=\sqrt{\varrho^2\Delta/\Sigma}$, $\beta^\varphi=-2ar/\Sigma$, $\gamma_{\varphi\varphi}=\Sigma\sin^2\theta/\varrho^2$, $\gamma_{rr}=\varrho^2/\Delta$, and $\gamma_{\theta\theta}=\varrho^2$. 
Here, we have defined $\varrho^2=r^2+a^2\cos^2\theta$, $\Delta=r^2+a^2-2r$, and $\Sigma=(r^2+a^2)^2-a^2\Delta\sin^2\theta$.
The fiducial observers (FIDOs), whose world lines are perpendicular to the hypersurface of $t= {\rm const.}$, are described by the coordinate four-velocity $n^\mu = (1/\alpha, -\beta^\varphi/\alpha, 0, 0)$. FIDOs rotate with coordinate angular velocity $\Omega = d\varphi/dt = -\beta^\varphi$. Maxwell equations in this $3+1$ formalism can be reduced to \citep{Kom04a}
\begin{equation}
    \partial_t \mathbf{B} + \nabla \times \mathbf{E} = 0, ~~~
    -\partial_t \mathbf{D} + \nabla \times \mathbf{H} = 4\pi \mathbf{J},
\end{equation}
$\nabla \cdot \mathbf{B} = 0$, and $\nabla \cdot \mathbf{D} = 4\pi \rho$. $\mathbf{D}$ and $\mathbf{B}$ are the electric and magnetic fields measured by FIDOs, while $\mathbf{E}$ and $\mathbf{H}$ are those in the coordinate basis, and they have relations
\begin{equation}
 \mathbf{E} =\alpha\mathbf{D}+\bm{\beta}\times\mathbf{B}, ~~~ 
 \mathbf{H} =\alpha\mathbf{B}-\bm{\beta}\times\mathbf{D}.
\end{equation}
In the following, we estimate the ratio of $\hat{B}_r = \sqrt{\gamma_{rr}} B^r$ and $\hat{B}_\varphi = B_\varphi/\sqrt{\gamma_{\varphi\varphi}}$, which are the radial and toroidal magnetic field comoponents with respect of the orthonormal basis carried by FIDOs.

\begin{figure}
   \begin{center}
    \includegraphics[width=\linewidth,pagebox=cropbox]{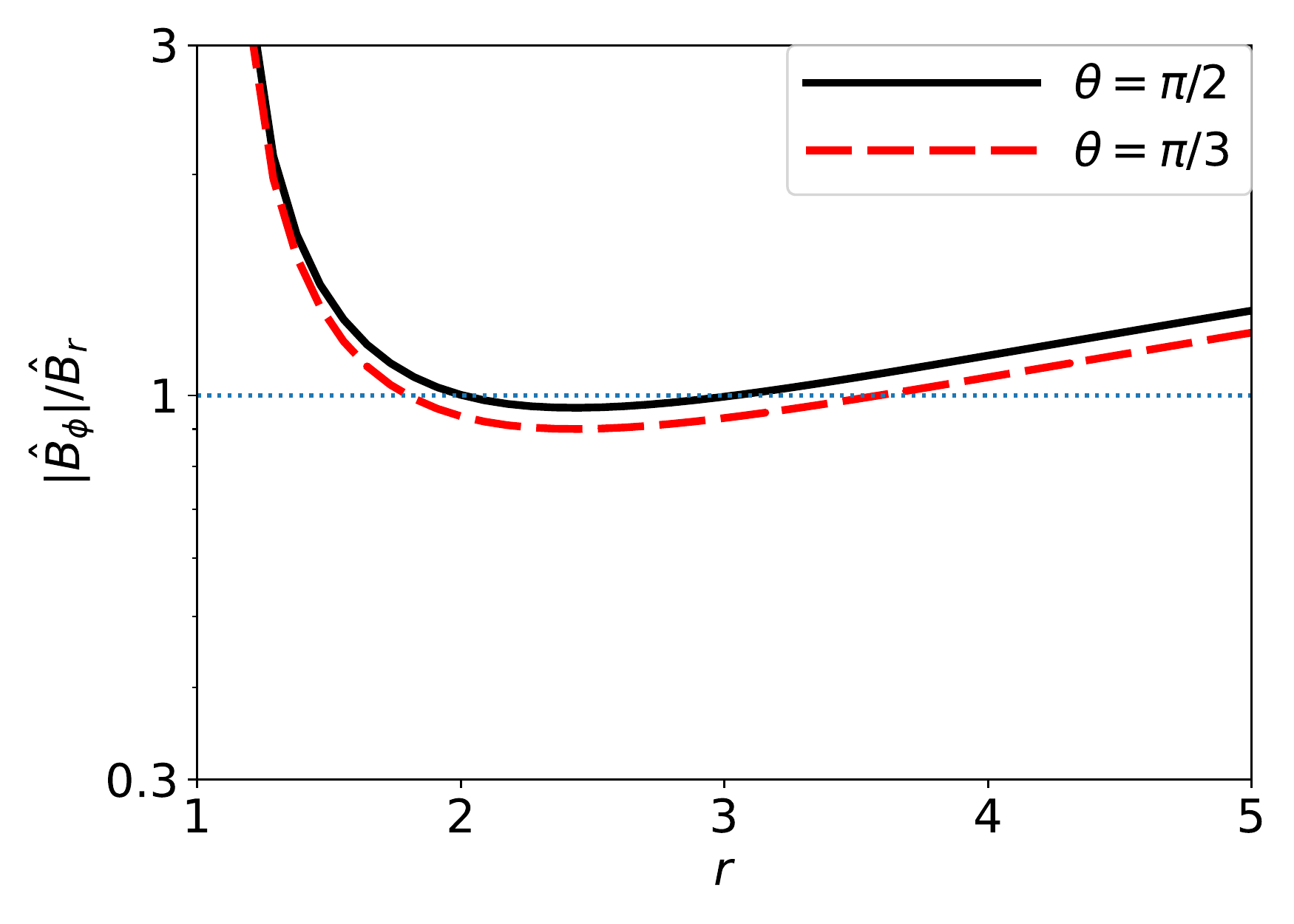}
    \caption{The ratio of toroidal to poloidal magnetic fields measured by FIDOs in the Boyer-Lindquist coordinates as a function of the distance from the BH horizon.}
    \label{fig:BphiBr}
   \end{center}
  \end{figure}

The high-resolution GRMHD simulation \citep{2022ApJ...924L..32R} shows that the accretion disk in the MAD state is so thin that the poloidal magnetic field in the BH magnetosphere has a monopole shape at $r \lesssim 5r_g$, i.e., $\sqrt{\gamma}B^r = {\rm const.}$ The force-free condition is a good approximation between the outer and inner light surfaces in the BH magnetosphere. Then, for the steady, axisymmetric state, the above Maxwell equations lead to $\mathbf{B}\cdot \nabla H_\varphi = 0$, i.e., $H_\varphi = \alpha B_\varphi = {\rm const.}$ along a magnetic field line \citep{Kom04a,TT16a}. Therefore, we have
\begin{equation}
    \frac{|\hat{B}_\varphi|}{\hat{B}_r} \propto \frac{\sqrt{\gamma_{\theta\theta}}}{\alpha}
\end{equation}
for a fixed $\theta$. Analyses of the special relativistic MHD outflows with high $\sigma$ \citep{2010mfca.book.....B,2013PTEP.2013h3E02T} indicate that
\begin{equation}
    \hat{B}_\varphi \sim -\hat{B}_r, ~~{\rm at} ~~ r \sim r_{\rm ols},
\end{equation} 
where $r_{\rm ols}$ means the radius of the outer light surface. Then we can plot $|\hat{B}_\varphi|/\hat{B}_r$ as shown in Figure \ref{fig:BphiBr}. Here we have assumed $a \sim 1$. The outer and inner light surfaces are located at $r \simeq 1.4$ and $3.0$, respectively, for $\theta=\pi/2$. In the region between the two light surfaces, which we are interested in, one has $|\hat{B}_\varphi|\sim\hat{B}_r$, and the magnetic field strength at the reconnection point can be estimated by Equation (\ref{eq:Brec}).

\section{Lepton loading mechanisms}\label{sec:massloading}

Several mechanisms of lepton loading into AGN jets have been studied.
Pair production by Comptonized MeV photons from hot accretion flows have been discussed for a long time \citep[e.g.,][]{2011ApJ...730..123L,2011ApJ...735....9M,2021ApJ...907...73W}, and this process may achieve $\kappa_\pm\sim 1-100$ for parameters of M87.  Hot accretion flows in the MAD state may efficiently accelerate non-thermal particles by magnetic reconnection, which leads to GeV gamma-ray production \citep{2020ApJ...905..178K,2022arXiv220509565K}. Since optical depth for $\gamma\gamma$ interactions for GeV gamma-rays is higher than that for MeV gamma-rays, this process can enhance the pair production rate at the magnetosphere, compared to the case only considering MeV gamma-rays. These mechanisms can efficiently create $e^+e^-$ pairs when accretion rate is high. However, the estimated multiplicity is $\sim100-10^3$, which is still lower than the required multiplicity estimated by radio observations ($\kappa_\pm\gtrsim10^5$; see e.g., \citealt{2015ApJ...803...30K}). 

Non-thermal hadronic particles, such as protons, neutrons, and atomic nuclei, may be loaded if hot accretion flows accelerate cosmic rays \citep{tt12,ktt14,kmt15}, but with the parameter set for M87, this process cannot achieve the multiplicity required by the radio observations.

Particle acceleration in the vacuum gap is also discussed as a lepton loading mechanism. \cite{2015ApJ...809...97B} argued that the multiplicity around the gap can be as high as $\sim100$, and it may further be enhanced to $\kappa_\pm\sim10^5$ at the far downstream of the jet. However, GR PIC simulations by \cite{2020ApJ...902...80K} found that the pair multiplicity around the gap is $\kappa_\pm\sim1$, and they also showed that $\kappa_\pm$ may not be so high even at the far downstream.  

Compared to these mechanisms, magnetic reconnection in BH magnetosphere can achieve many orders of magnitude higher multiplicity because of the efficient, local conversion of magnetic energy into MeV photon energy.
In our scenario, high-density $e^+ e^-$ pairs are intermittently loaded into the limited region in BH magnetosphere, and thus, the quiescent state or the other part of the BH magnetosphere could have the vacuum gap which may produce very-high-energy gamma-rays \citep{2022ApJ...924...28K}.

\section{Particle Acceleration by Relativistic Magnetic Reconnections} \label{sec:acceleration}

Many PIC simulations show that relativistic reconnection very efficiently accelerates non-thermal particles, although the details of the mechanism are still controversial.  
As a pioneering work, \cite{2001ApJ...562L..63Z} showed that using 2D PIC simulations, the reconnection electric field generated at the X-point accelerates particles very rapidly with a timescale of $t_{\rm acc}\approx\gamma_e m_e c/(eB\beta_{\rm rec} )$. We use this timescale as the fiducial model, as this is a simple and robust mechanism.

PIC simulations with larger box sizes and longer durations have been available since 2010s. \cite{2014PhRvL.113o5005G} revealed that the X-point acceleration is sub-dominant, and Fermi-like acceleration is essential using a set of 2D PIC simulations. The acceleration timescale can be represented by $t_{\rm acc,G14}\approx (\sigma_B/10)^{-1/2} l_{\rm rec}/c$, where $l_{\rm rec}$ is the size of the reconnecting magnetic field. \cite{2021ApJ...919..111G} reached similar conclusions with 3D simulations with various parameter sets.  However, this acceleration timescale is not applicable to our scenario. For the Fermi-type acceleration to work effectively, the gyration timescale, $t_L\approx E_e/(eBc)$, needs to be shorter than the acceleration timescale, whereas $t_{\rm acc,G14}\ll t_L$ is satisfied for typical parameters used in our calculations. More recently, \cite{2022PhRvL.128n5102S} revealed that the electric fields at the X-point play an essential role in the initial phase of the particle acceleration, and thus, our treatment focusing on the X-point acceleration is reasonable.

\cite{2021ApJ...922..261Z} argued that the particle acceleration mainly occurs by the electric fields at X-points. In 3D simulations, the reconnection dynamics is turbulent. Energetic particles are not trapped in magnetic islands, and some of them continuously gain energy from the X-point electric fields. This mechanism works for particles of $\gamma_e \gtrsim 3\sigma_B$, but the lower-energy particles should be confined in the magnetic islands even in 3D simulations. Since our scenario focuses on particles of $\gamma_e\ll \sigma_B$, their model is inappropriate to our scenario.

\cite{2018PhRvL.121y5101C,2019ApJ...886..122C} showed that the reconnection acceleration by turbulent medium leads to anisotropic pitch angle distribution. This is caused by the strong guide fields at the reconnection points due to turbulent nature. In this case, the synchrotron emission is suppressed, and particles may be accelerated to higher energies than that for the isotropic distribution. Then, we expect gamma-rays of energies higher than that given in Equation (\ref{eq:Egammax}). Nevertheless, \cite{2022ApJ...924L..32R} found that the magnetic reconnection in BH magnetoshpere proceeds with almost no guide field, which is assumed in our scenario. We speculate that some different initial magnetic-field configuration may lead to magnetic reconnnection with a strong guide field, as seen in \citet{BOP18a}, which may lead to production of higher energy gamma-rays.

\citet{2013ApJ...770..147C} performed a set of reconnection simulations with the effect of synchrotron radiation feedback. They found that some fraction of electrons can achieve energies higher than that given by Equation (\ref{eq:gamemax}), because the magnetic field at the X-point is much weaker than the upstream plasma. These high-energy electrons lead to the photon spectrum with a high-energy tail above the burn-off limit ($\sim160$ MeV). However, the emerging photons above the burn-off limit is sub-dominant in terms of the luminosity and the number density. Also, the high-energy photons above 160 MeV are strongly beamed to the direction of the reconnecting magnetic field lines, which will significantly reduce the pair production rate. For these reasons, we concentrate on the pair production and emerging spectra for the lower energy photons in this Letter. 

\cite{2019ApJ...877...53H} performed a set of reconnection simulations taking into account the synchrotron energy loss and $e^+e^-$ pair production. Their simulations also exhibit production of gamma-rays of energies higher than that given in Equation (\ref{eq:Egammax}). However, they performed only 2D simulations, and it is unclear how 3D effects will modify the emerging photon spectra. 

\citet{2021MNRAS.507.5625S,2022arXiv220302856S} performed 2D PIC simulations taking account of the Compton cooling. Their results show that spectra for non-thermal particles are steeper for the cases with stronger cooling, which might indicate that the gamma-ray emission could be suppressed if the cooling is efficient. Nevertheless, the Compton cooling is different from the synchrotron cooling: The Compton cooling is efficient for electrons of any pitch angles and positions, whereas the synchrotron emission more efficiently cools electrons of larger pitch angles and at the regions of stronger magnetic fields.

\section{Pair production rate and  Multiplicity}\label{sec:multiplicity}
The pair production rate is estimated to be
\begin{align}
 \dot{N}_{\gamma\gamma}=2\int n_{\gamma_1}n_{\gamma_2}\sigma_{\gamma\gamma}cdV,
\end{align}
where $n_{\gamma_1}$ and $n_{\gamma_2}$ are the number density of interacting photons, $\sigma_{\gamma\gamma}$ is the cross-section, and $dV$ is the volume of the interacting region. 
Here, we consider a uniform emission region and $\gamma\gamma$ interactions inside the sphere of size $l_{\rm rec}$.
The photons of $E_{\gamma,\rm max}$ mainly interact with the photons of $E_{\gamma_2} \approx (2m_ec^2)^2/E_{\gamma,\rm max} $, whose number density can be
\begin{align}
 n_{\gamma_2}\approx 
 \frac{ L_{\rm rec} (E_{\gamma_2}/E_{\gamma,\rm max})^{1/2}}
 {4\pi l_{\rm rec}^2cE_{\gamma_2}}\approx \frac{L_{\rm rec}}{8\pi l_{\rm rec}^2 m_ec^3}.
\end{align}
The optical depth for the $\gamma\gamma$ interaction is then given by Equation (\ref{eq:taugam}).
Then the pair production rate can be approximated to be
\begin{align}
 \dot{N}_{\gamma\gamma}\approx \frac{L_{\rm rec}\tau_{\gamma\gamma}}{E_{\gamma,\rm max}}.\label{eq:dotNgam}
\end{align}
The number density of the $e^+e^-$ pairs can be written as \citep[e.g.,][]{2020ApJ...905..178K}
\begin{equation}
 n_\pm \approx \frac{3\dot{N}_{\gamma\gamma}}{l_{\rm rec}^2c}.\label{eq:npm}
\end{equation}
Using Equations (\ref{eq:sigBgj}), (\ref{eq:Lrec}), (\ref{eq:taugam}), (\ref{eq:dotNgam}), and (\ref{eq:npm}), we obtain Equation (\ref{eq:kappapm}).

We can approximately write the injection spectrum of the $e^+e^-$ pairs as $\dot{N}_{E_\pm,\rm inj}\propto E_\gamma^{-1}$. Suppose that the photon spectrum is a power-law form, $E_\gamma n_{E_\gamma}\approx n_0(E_\gamma/E_{\gamma,\rm max})^s$, where $n_0$ is a normalization factor. A photon of $E_{\gamma,1}$ interact with photons of $E_{\gamma,2}\approx (2m_ec^2)^2/E_{\gamma,1}$. The number density of these photons are given by 
\begin{align}
n_{\gamma,1}&\approx n_0 \left(\frac{E_{\gamma,1}}{E_{\gamma,\rm max}}\right)^s,\\
n_{\gamma,2}&\approx n_0 \left(\frac{4m_e^2c^4}{E_{\gamma,1}E_{\gamma,\rm max}}\right)^s.
\end{align}
We approximate $\sigma_{\gamma\gamma}$ as a delta function, and the mean energy of the $e^+e^-$ pair produced by photons of $E_\gamma$ is $E_\pm \approx E_\gamma/2$. Then, the spectrum of the $e^+e^-$ pairs can be given by 
\begin{align}
 E_\pm \dot{N}_{E_\pm,\rm inj}&\approx 2c f_{\gamma\gamma}\sigma_T n_{\gamma,1} n_{\gamma,2}V\nonumber\\
 &\approx 2cf_{\gamma\gamma}\sigma_Tn_0^2V\left(\frac{2m_ec^2}{E_{\gamma,\rm max}}\right)^{2s}
\end{align}
where $V$ is the volume of the injection region.
This does not depend on $E_{\gamma,1}$, so that
we can write $\dot{N}_{\pm,\rm inj}\propto E_\pm^{-1}$.

\section{Energy loss in the lepton plasma}\label{sec:SSAheating}
The SSA heating timescale is written as  \citep[e.g.,][]{2011ApJ...739..103A}
\begin{align}
     t_{\rm ssa}=\frac{\gamma_\pm m_ec^2}{\int dE_\gamma E_\gamma n_{E_\gamma}\sigma_{\rm ssa}c},
\end{align}
where $n_{E_\gamma}$ is the differential photon number density, $\gamma_\pm=E_\pm/(m_ec^2)$, and $\sigma_{\rm ssa}$ is the SSA cross section \citep{1991MNRAS.252..313G}.
With the hard lepton injection spectrum of $dN/dE_\pm\propto E_\pm^{-1}$, the synchrotron spectrum is dominated by the photons produced by the highest energy leptons. 
The SSA heating rate by the lowest energy photons of
$\nu_{\rm min}=eB_{\rm rec}/(2\pi m_ec\gamma_{\pm,\rm max})$ is dominant, 
which allows us to write 
\begin{align}
 &\int dE_\gamma E_\gamma n_{E_\gamma}\sigma_{\rm ssa}\nonumber\\
&\approx \frac{\dot{N}_{\gamma\gamma}E_{\pm,\rm max}}{4\pi l_{\rm rec}^2c}{\rm min}\left(\frac{t_{\rm dyn}}{t_{\pm,\rm syn}},~1\right) \left(\frac{\nu_{\rm min}}{\nu_{\rm max}}\right)^{4/3}\sigma_{\rm ssa0},
\end{align}
where
\begin{equation}
\sigma_{\rm ssa0}\approx\frac{2^{11/3}3^{7/6}\pi^2\Gamma^2(4/3)}{5}\frac{e}{B_{\rm rec}},
\end{equation}
$\nu_{\rm min}=eB_{\rm rec}/(2\pi \gamma_{\pm,\rm max}m_ec)$ and $\nu_{\rm max}=eB_{\rm rec}\gamma_{\pm,\rm max}^2/(2\pi m_ec)$, $\Gamma(x)$ is the Gamma function, 
$t_{\rm dyn}=R_{\rm rec}/c$ is the dynamical timescale, and $t_{\pm,\rm syn}$ is the synchrotron cooling timescale. 
For our reference parameter set, we obtain
\begin{align}
    t_{\rm ssa}&\approx \frac{4\pi l_{\rm rec}^2}{\dot{N}_{\gamma\gamma}\sigma_{\rm ssa0}}\left(\frac{\nu_{\rm min}}{\nu_{\rm max}}\right)^{-4/3}{\rm max}\left(\frac{t_{\pm,\rm syn}}{t_{\rm dyn}},~1\right)\nonumber \\
&\simeq4.8 M_9^{1/2}\dot{m}_{-4}^{-3/2}f_l^{-1}\Phi_{\rm rec,1.2}^{-3}\beta_{\rm rec,-1}^3 \rm~s,
\end{align}
\begin{align}
t_{\pm, \rm syn}&\simeq 3.3\times10^2 M_9\dot{m}_{-4}^{-1}\Phi_{\rm rec,1.2}^{-2}\beta_{\rm rec,-1}^{-1}\rm~s,\\
t_{\rm dyn}&\simeq9.8\times10^3M_9\rm~s. 
\end{align}
Since $t_{\rm ssa}<t_{\pm,\rm syn}$ for the entire energy range, the $e^+e^-$ plasma should be thermalized without energy loss. M87 satisfies this condition. 

For the cases with $t_{\rm dyn}<t_{\pm,\rm syn}$, which is satisfied in Sgr A*, the lepton plasma does not efficiently lose energy by radiation. 
Then, half of the injected plasma could move outward, and we can estimate the radio luminosity by 
\begin{equation}
\mathcal{E}_{\rm blb}=\frac{1}{2}\dot{N}_{\gamma\gamma}T_{\rm dur}E_{\pm,\rm max}. 
\end{equation}
The other half of the injected plasma will fall to the SMBH. We consider that the SSA heating is ineffective in this case, regardless of the value of $t_{\rm ssa}$.

For the cases with $t_{\rm ssa}>t_{\pm,\rm syn}$ and $t_{\rm dyn}>t_{\pm,\rm syn}$, the lepton plasma lose energy by radiation. In this case, 
we use 
\begin{equation}
\mathcal{E}_{\rm blb}=\frac{1}{4}\dot{N}_{\gamma\gamma}T_{\rm dur}E_{\pm,\rm max}, 
\end{equation}
where we assume that only the momentum parallel to the magnetic field is left after synchrotron cooling. Here, we consider that half of the injected plasma move outward.


\bibliography{ssk}
\bibliographystyle{aasjournal}



\end{document}